\documentstyle[preprint,aps]{revtex}
\begin{document}
\normalsize
\title { Production of Strange Clusters and Strange Matter in
Nucleus-Nucleus Collisions at the AGS}
\author{P. Braun-Munzinger and J. Stachel$^*$}
\address {\em Department of Physics \\
       State University of New York at Stony Brook \\
       Stony Brook, \, New York 11794 -- 3800 }
\date{\today}
\maketitle
\begin{abstract}
Production probabilities for strange clusters and strange matter in
Au+Au collisions at AGS energy are obtained in the thermal fireball
model. The only parameters of the model, the baryon chemical potential
and temperature, were determined from a description of the rather
complete set of hadron yields from Si+nucleus collisions at the AGS. For
the production of light nuclear fragments and
strange clusters the results are similar to recent coalescence model
calculations. Strange matter production with baryon number larger than
10 is predicted to be much smaller than any current experimental sensitivities.
\end{abstract}
\narrowtext

In a recent study \cite{plb_eq} of thermal
equilibration and expansion in nucleus-nucleus collisions at AGS
energies we have shown that a consistent description of the production
of hadrons in Si-nucleus collisions can be obtained  under the assumption that
strangeness is completely equilibrated. Here we explore the predictions
of this thermal model with respect to the formation of light nuclear
fragments and strange clusters and focus, in particular, on  production
probabilities of strange matter in Au+Au collisions at AGS energy.

The possibility that matter composed of a roughly equal mixture of up-,
down-, and strange quarks is metastable or even stable has been a
subject of intense recent investigations. An introduction and comprehensive
survey of experimental and theoretical considerations can be found in
\cite{aarhus}. A number of experimental searches are currently underway
to test the possible existence and (meta)stability of this new phase of
matter. Since strange quark matter has a very large strangeness to baryon
number ratio \cite{berger}  of $|S|/B \approx 0.8$,  most of the
searches focus on
ultrarelativistic nucleus-nucleus collisions, where significant numbers
of strange particles can be produced per interaction. In fact, at AGS
energies,  strangeness production as measured by the probability
$\lambda = \frac{2 s \bar s} {u \bar u +d \bar d}$ is enhanced compared
to that found in nucleon-nucleon collisions by about a factor of two
\cite{sy,pbms}.

In the present
investigation we use a fireball model based on the assumption of thermal
and chemical equilibrium to make predictions for strange matter
production probabilities in such collisions.
The success of this model reported in \cite{plb_eq} suggests  that, at
AGS energies,
thermal and chemical equilibrium are, indeed, achieved. Within such a
thermal model, particle production does not anymore depend on specific
(and, in general, unknown) cross sections but rather is governed by
conservation laws and the baryon chemical potential and temperature of
the system at freeze-out, as well as by the mass and quantum numbers of
the particle to be produced. The success of the model for strangeness
production in general implies that the present estimates should provide
a baseline for strange matter production from hadronic matter at
freeze-out. Whether quark-gluon plasma formation and subsequent
hadronization can lead to  strange matter production which is enhanced
beyond the present estimate which is  based on chemical equilibration is
unclear at present.

As shown in \cite{plb_eq} the available data on hadron yields in central
Si-nucleus collisions at 14.5 GeV/c per nucleon can be quantitatively
described in a fireball model under the assumption of thermal and
chemical equilibrium. In this model, the particle number densities are
given as integrals over particle momentum $p$:
\begin{equation}
 \rho_i^0 = \frac{g_i}{2\pi^2} \int_{0}^{\infty} \frac{p^2 dp}{{\rm
exp}[(E_i-\mu_b B_i - \mu_s S_i)/T] \pm 1}
\end{equation}
where $g_i$ is the spin-isospin degeneracy of particle {\it i}, $E_i$,
$B_i$ and $S_i$ are its total energy in the local restframe, baryon
number and strangeness, and $\mu_b$ and $\mu_s$ are the baryon and
strangeness chemical potentials ($\hbar$=c=1). As
in \cite{plb_eq} we employ finite and excluded volume
corrections. Analysis of the Si-nucleus data yields a value of $\mu_b$ =
0.54 GeV and a temperature range of $0.12 < T < 0.14$ GeV.  For a given
temperature and baryon chemical potential the strangeness chemical
potential $\mu_s$ is fixed by strangeness conservation. In particular,
for $T$ = 0.120 and 0.140 GeV one obtains $\mu_s$ = 0.108 and 0.135 GeV.
This thermal approach leads to a good description of all strange
particle yields measured so far at AGS energy \cite{plb_eq}.

In Table\ \ref{ratios} we show that the model for Si+Pb collisions also very
well reproduces the measured d/p and \=d/\=p ratios, without explicit
reference to the underlying
coalescence mechanism which presumably governs the production of such
loosely bound clusters. To explore the thermal model
predictions for more complex clusters we have also calculated production
probabilities in central Au-Au collisions for ordinary and
(multi)-strange clusters
recently discussed in ref. \cite{dover1}. The results are shown in
Table\ \ref{probab}. To compute absolute production
probabilities one has, of course, to specify a freeze-out volume. We
have chosen this volume such that the total pion multiplicity (charged
and neutral) is 450 for central Au-Au collisions, in accord with
preliminary data from experiment E866 \cite{gonin}. The corresponding
freeze-out volume is 2600 and 5000 $fm^3$, for $T=0.14$ and $T=0.12$
GeV, respectively, quite reasonable in view of the recent estimate for
the freeze-out volume from Si+Pb collisions \cite{E814_hbt}. An alternative
possibility would have been to fix the number of baryons in the fireball
to 394, leading to very similar results.
As shown in Table\ \ref{probab} the thermal
model predicts deuteron, t, $^3$He, and $\alpha$ particle production
in close agreement with
results from the coalescence model. This latter fact is particularly
relevant in connection with
thermal model predictions of strange matter production since strange
matter is predicted \cite{chin} to have similar central density as
$^4$He.
The results for
strange cluster production are also compiled in Table\
\ref{probab}. They agree rather  well with the coalescence model
calculations reported in \cite{dover1}, lending support to the
conclusion that such clusters, should they exist as (quasi)bound states,
should be observable in Au+Au collisions at AGS energy. Close inspection
of Table\ \ref{probab} reveals that the coalescence model of
\cite{dover1} predicts somewhat larger production probabilities compared
to the thermal model for
multistrange clusters with baryon number larger than 4. The origin may
be the particular choice of the coalescence parameters \cite{dover2}.
We stress again
that only the baryon number, mass, spin and strangeness of the cluster
enters the thermal model calculation here and the results are
independent of  particular cluster wave functions and
coalescence parameters.

To compute production probabilities for strange matter clusters (or
strangelets) we use
the Berger and Jaffe \cite{berger} mass formula to determine the most
stable charge and strangeness of a cluster with given baryon number
B. For the energy per baryon of each cluster we use the recent mass formula
of \cite{madsen}, where curvature and surface effects are properly
calculated for small clusters. Using the total energy, baryon number,
and strangeness we are now in a position to estimate the strangelet
production in central Au-Au collisions at AGS energy. The results are
again listed in Table\
\ref{probab}. Even for the
$_1^{10}$St$^{-8}$ strangelet with baryon number 10, charge number 1, and
strangeness $S = - 8$,  which is predicted to be unstable against
strong decays   \cite{madsen}, the
production
probability is less than $10^{-12}$ per central Au-Au collision. For
each increase in baryon number by one, one has to pay a penalty factor
of the order of $ \exp(-\Delta m/T)\exp((\mu_b-\mu_s)/T)$, where $\Delta m$ is
the corresponding increase in strangelet mass. This amounts to about a
factor  50 and 70 for $T = 0.14$ and $T = 0.12$ GeV, respectively. Note that,
in our model, even ordinary nuclei with baryon number larger than 10 are
only produced at minute rates in central Au+Au collisions: for $^{12}$C
production at $T = 0.14$ GeV we predict a production probability of $5.4
\cdot 10^{-11}$ per central Au-Au collision. This should be compared
with a production probability of $ 1.7 \cdot 10^{-15}$ for a strangelet
with baryon number $B = 12$ (see Table \ \ref{probab}).
Our calculations produce results which are many orders of
magnitude  smaller than the
predictions by \cite{crawford}, where a specific quark matter production
scenario is invoked to estimate strange matter production. Furthermore,
the authors of \cite{crawford} used a much more optimistic mass
formula. In the present model, a reduction in strangelet mass by an
amount $\delta
m$, keeping all other parameters fixed, increases the production
probability by about $\exp(\delta m/T)$.  The uncertainty in strangelet mass
consequently leads to a substantial uncertainty in the production
probability. However, even the most optimistic scenario for strangelet
mass ({\it i.e.} assuming a bulk binding parameter of 0.88 GeV as in
\cite{crawford}) increases the production probability for {\it e.g.} the
$_1^{12}$St$^{-9}$ strangelet from $1.7 \cdot 10^{-15}$ to $2.7 \cdot
10^{-13}$, still  below the sensitivity of any existing or proposed
experiment.

The thermal model employed here describes all available data on strange
and nonstrange hadron production in Si-nucleus collisions at AGS energy,
without need to discuss
specific reaction mechanisms. Since thermal and chemical equilibrium are
the major ingredients of our model, all indications are that it will be
applicable also for  Au+Au collisions. It should be investigated whether models
such as the strangeness distillation
scenario proposed in \cite{cgreiner} and invoked in \cite{crawford},
where locally more  strangeness is produced than in a fully equilibrated hadron
gas at freeze-out, correctly predict  present data on
strange hadron production. We note that, for the quark gluon plasma, the
strangenes suppression factor $\lambda$ is indeed large when calculated
in a thermal model, but decreases significantly during hadronization
\cite{pbms}.

In conclusion, the fireball model based on thermal and chemical
equilibrium describes cluster formation well, where measured. It gives results
similar in magnitude to the predictions of the coalescence model
developed recently \cite{dover1} to estimate production probabilities
for light nuclear fragments (p, d, t, $\alpha$ ...) and for
for strange hadronic clusters (such as the H dibaryon) in Au-Au
collisions at the AGS. Predicted
yields for production of strange matter with baryon number larger than
10 are well below current experimental sensitivities.

$^*$ This work was supported in part by the National Science Foundation.

\begin{table}
\caption{Particle ratios calculated in a thermal model for two different
temperatures, baryon chemical potential $\mu_b$= 0.54 GeV and
strangeness chemical potential $\mu_s$ such that overall strangeness is
conserved, in comparison to experimental data (with statistical errors
in parentheses) for central collisions of 14.6 A GeV/c Si + Au.
\label{ratios}}
\begin{tabular}{||c|ll|llc||}
Particles & \multicolumn{2}{c|} {Thermal Model} &
\multicolumn{3}{c||}{Data}\\
 & $T$=.120 GeV & $T$=.140 GeV & exp. ratio & rapidity & ref.\\ \hline
d/(p+n)     & 4.3 $\cdot$ 10$^{-2}$ & 5.8 $\cdot$ 10$^{-2}$ & 3.0(3)
$\cdot$
10$^{-2}$ & 0.4 - 1.6 & \cite{802prc} \\
 \=d/\=p & 1.1 $\cdot$ 10$^{-5}$ & 4.7 $\cdot$ 10$^{-5}$ &
1.0(5)$\cdot$ 10$^{-5}$ & 2.0 & \cite{858dbar}\\
\end{tabular}
 \end{table}

\begin{table}
\caption{Produced number of nonstrange and strange clusters and
of strange quark
matter  per  central Au+Au collisions at AGS energy, calculated in a
thermal model for two different
temperatures, baryon chemical potential $\mu_b$= 0.54 GeV and
strangeness chemical potential $\mu_s$ such that overall strangeness is
conserved. The coalescence model predictions are from Table 2 of [6].
\label{probab}}
\begin{tabular}{||l|ll|l||}
Particles & \multicolumn{2}{c|} {Thermal Model} & {Coalescence Model}\\ \hline
 & $T$=.120 GeV & $T$=.140 GeV  & \\ \hline
d & 15 & 19 & 11.7  \\
t+$^3$He & 1.5 & 3.0 & 0.8  \\
$\alpha$ & 0.02 & 0.067 & 0.018 \\
$H_0$       & 0.09 & 0.15 & 0.07  \\
$_{\Lambda \Lambda}^5$H & 3.5 $\cdot 10^{-5}$ & 2.3 $\cdot 10^{-4}$ &
4$\cdot 10^{-4}$ \\
$_{\Lambda \Lambda}^6$He & 7.2 $\cdot 10^{-7}$ & 7.6 $\cdot 10^{-6}$ &
 1.6$\cdot 10^{-5}$ \\
$_{\Xi^0\Lambda \Lambda}^7$He & 4.0 $\cdot 10^{-10}$ & 9.6 $\cdot
10^{-9}$  & 4 $\cdot 10^{-8}$  \\ \hline
$^{10}_{1}$St$^{-8}$ &  1.6 $\cdot 10^{-14}$ & 7.3 $\cdot 10^{-13}$ & \\
$^{12}_1$St$^{-9}$ &  1.6 $\cdot 10^{-17}$ & 1.7 $\cdot 10^{-15}$  &  \\
$^{14}_1$St$^{-11}$ & 6.2 $\cdot 10^{-21}$ & 1.4 $\cdot 10^{-18}$  & \\
$^{16}_2$St$^{-13}$ & 2.4 $\cdot 10^{-24}$ & 1.2 $\cdot 10^{-21}$   & \\
$^{20}_2$St$^{-16}$ & 9.6 $\cdot 10^{-31}$ & 2.3 $\cdot 10^{-27}$   & \\
\end{tabular}
\end{table}

\end{document}